\documentclass[useAMS,usenatbib]{mn2e}
\usepackage[dvips]{graphicx}
\usepackage{epsfig}
\usepackage{rotating}
\usepackage{natbib}
\usepackage{color}
\usepackage{mathrsfs}
\usepackage{amssymb}
\usepackage[caption=false]{subfig}

\newcommand{\Msun}{\, {\rm M_{\odot}}}

\title[Morphologies and colours of the EAGLE galaxy population]{The relation between galaxy morphology and colour in the EAGLE simulation}
\author[C.A.~Correa et al.]
 {Camila A.~Correa$^{1}$, Joop~Schaye$^1$, Bart Clauwens$^{1,2}$, Richard G. Bower$^3$, 
 \newauthor Robert A. Crain$^4$, Matthieu Schaller$^{3}$, Tom Theuns$^3$ and Adrien C. R. Thob$^4$\\
 $^1$ Leiden Observatory, Leiden University, P.O. Box 9513, 2300 RA Leiden, The Netherlands\\
 $^2$ Instituut-Lorentz for Theoretical Physics, Leiden University, 2333 CA Leiden, The Netherlands\\
 $^3$ Institute for Computational Cosmology, Physics Department, University of Durham, South Road, Durham DH1 3LE, UK\\
 $^4$ Astrophysics Research Institute, Liverpool John Moores University, 146 Brownlow Hill, Liverpool L3 5RF, UK\\
 }

\date{\today}

\pagerange{\pageref{firstpage}--\pageref{lastpage}} \pubyear{2013}

\def\LaTeX{L\kern-.36em\raise.3ex\hbox{a}\kern-.15em
    T\kern-.1667em\lower.7ex\hbox{E}\kern-.125emX}

\begin{document}
\maketitle

\begin{abstract}
We investigate the relation between kinematic morphology, intrinsic colour and stellar mass of galaxies in the EAGLE cosmological hydrodynamical simulation. We calculate the intrinsic $u-r$ colours and measure the fraction of kinetic energy invested in ordered corotation of 3562 galaxies at $z=0$ with stellar masses larger than $10^{10}\Msun$. We perform a visual inspection of {\it{gri}}-composite images and find that our kinematic morphology correlates strongly with visual morphology. EAGLE produces a galaxy population for which morphology is tightly correlated with the location in the colour-mass diagram, with the red sequence mostly populated by elliptical galaxies and the blue cloud by disc galaxies. Satellite galaxies are more likely to be on the red sequence than centrals, and for satellites the red sequence is morphologically more diverse. These results show that the connection between mass, intrinsic colour and morphology arises from galaxy formation models that reproduce the observed galaxy mass function and sizes.
\end{abstract}


\section{Introduction}

The morphology of galaxies is typically characterised either by extensive visual inspection (e.g. Galaxy Zoo project, \citealt{Lintott11}) or through the stellar kinematics (e.g. \citealt{Emsellem07}). Both methods can be used to determine whether a galaxy is bulge- or disc-dominated or whether it appears disturbed (for example experiencing a merger). It is well established that disc-dominated galaxies tend to be rotationally supported, spheroidal galaxies dispersion supported, and that overall morphology strongly correlates with galaxy mass, with massive galaxies being mostly bulge-dominated (e.g. \citealt{Sandage78,Conselice06,Ilbert10,Bundy10}). It has also been shown that morphology correlates with galaxy colour (e.g. \citealt{Larson80,Strateva01,Baldry04}) and that galaxies can be divided into two well-defined distinct populations, namely `red-sequence' galaxies, that tend to be elliptical, bulge-dominated, older and redder than `blue-cloud' galaxies, which are disc-dominated and star-forming . 

Despite recent progress in numerical simulation predictions of galaxy morphologies (\citealt{Snyder15,Dubois16,Rodriguez-Gomez17,Bottrell17}) and increasing observational data (\citealt{Willett13,Haubler13}), the origin of the distribution of galaxy morphologies and its correlation with star-forming/quenching galaxies is still under debate. Cosmological simulations of galaxy formation aim to reproduce and provide an explanation for the origin of the global properties and scaling relations revealed by galaxy surveys. The EAGLE cosmological simulation suite (\citealt{Schaye15,Crain15}) reproduces many properties of the observed galaxy  population including the evolution of galaxy masses (\citealt{Furlong15}), sizes (\citealt{Furlong17}), star formation rates and colours (\citealt{Trayford15,Trayford16}), and black hole masses and AGN luminosities (\citealt{Rosas16,Bower17,McAlpine17}) with unprecedented accuracy. \citet{Trayford16} examined the origin of intrinsic colours of EAGLE galaxies, and found that while low-mass red-sequence galaxies are mostly satellites (indicative of environmental quenching), most high-mass red-sequence galaxies have relatively large black hole masses (indicative of internal quenching due to AGN feedback, see \citealt{Bower17}).

Here we present the first investigation of the relationship between morphology and intrinsic colour for the $z=0$ EAGLE galaxy population. The two main approaches to characterise a simulated galaxy's morphology are predicting the surface brightness and determining the bulge-to-disc ratio using a decomposition into S\'ersic profiles (e.g. \citealt{Snyder15}) or quantifying the rotational support based on stellar motions (e.g. \citealt{Abadi03,Sales12,Dubois16,Rodriguez-Gomez17}). Although the results of these methods correlate, the scatter is large and photometric decompositions tend to result in lower bulge-to-disc ratios (e.g. \citealt{Scannapieco10,Bottrell17}). Determining morphology from (mock) images is more challenging since it requires modeling the effects of colour gradients, extinction and scattering, projection and background subtraction. We leave such a photometric study of galaxy morphology, as well as a comparison to observations, for future work and focus on the correlation between kinematic morphology and colour. The outline of this letter is as follows. In Section 2 we briefly describe the simulation, the galaxy selection and the computation of galaxy colours and morphology, whose correlation is analysed in Section 3. We summarize our results in Section 4.

\section{Methodology}
\subsection{The EAGLE simulation}

The EAGLE suite (\citealt{Schaye15,Crain15}) consists of a series of cosmological, hydrodynamical simulations, run with a  modified version of GADGET 3 (\citealt{Springelb}), a $N$-Body Tree-PM smoothed particle hydrodynamics (SPH) code. Throughout this work we analyse the reference model (Ref) run in a cosmological volume of 100 comoving Mpc on a side with an initial gas and dark matter particle mass of $m_{\rm{g}}=1.8\times 10^{6}\Msun$, $m_{\rm{dm}}=9.7\times 10^{6}\Msun$, respectively, and a Plummer equivalent gravitational softening smaller than $\epsilon=0.7$ proper kpc. It assumes a $\Lambda$CDM cosmology with the parameters derived from {\it{Planck-1}} data (\citealt{Planck}). 

EAGLE uses the hydrodynamics solver ``Anarchy'' that adopts the pressure-entropy formulation described by \citet{Hopkins13}, an artificial viscosity switch as in \citet{Cullen10}, and an artificial conduction switch. The reference model includes radiative cooling and photo-heating (\citealt{Wiersma09a}), star formation (\citealt{Schaye08}), stellar evolution and mass loss (\citealt{Wiersma09b}), black hole growth (\citealt{Springel05,Rosas15}) and feedback from star formation and AGN (\citealt{DallaVecchia12}). The subgrid model for stellar feedback was calibrated to reproduce the observed $z=0$ galaxy mass function, the mass-size relation for discs.

Dark matter halos (and the self-bound substructures within them associated to galaxies) are identified using the Friends-of-Friends (FoF) and SUBFIND algorithms (\citealt{Dolag}). In each FoF halo, the `central' galaxy is the subhalo at the minimum of the potential, usually the most massive. The remaining galaxies within the FoF halo are its satellites. To determine the mass (and luminosity) of galaxies, we follow \citet{Schaye15} and calculate them in spherical apertures of 30 kpc. To prevent resolution effects, we select galaxies with stellar masses larger than $10^{10}\Msun$, which results in a sample of 3562 galaxies.

\begin{figure} 
\begin{center}
\vspace{-0.4cm}
\includegraphics[width=0.34\textwidth]{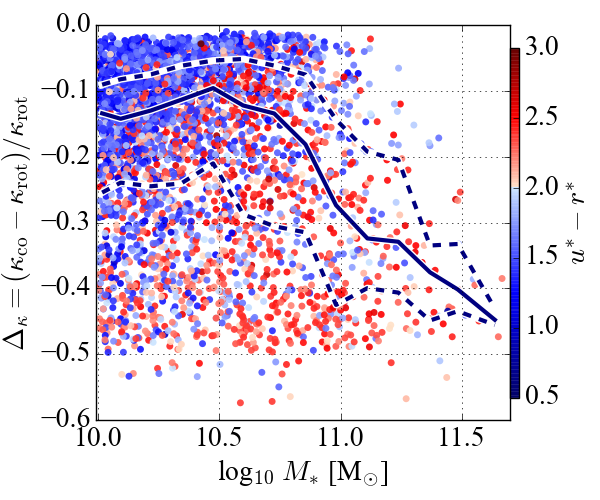}
\end{center}
\vspace{-0.4cm}
\caption{Scatter plot of the relative difference between $\kappa_{\rm{co}}$ and $\kappa_{\rm{rot}}$ as a function of stellar mass. Individual galaxies are plotted as points, coloured by intrinsic $u^{*}-r^{*}$ according to the colour bar on the right. The solid line shows the median relation and the dashed lines the 25th and 75th percentiles.}
\label{color_morpho_distribution_1}
\end{figure}

\subsection{Galaxy colour and morphology}

We use the galaxy colours computed by \citet{Trayford15}, who adopted the GALAXEV population synthesis model of \citet{Bruzual03}, that provide the spectral energy distribution (SED) per unit initial stellar mass of a simple stellar population (SSP) for a discrete grid of ages and metallicities. The SED of each stellar particle was computed by interpolating the GALAXEV tracks logarithmically in age and metallicity, and multiplying by the initial stellar mass. In EAGLE, each star particle represents an SSP characterised by the \citet{Chabrier03} IMF over the mass range [0.1,100]$\Msun$. The spectra were summed over all stars within a spherical aperture of 30 kpc and convolved with a filter response function. We use the intrinsic (i.e. rest-frame and dust-free) ${u^{*}-r^{*}}$ colours (with the $*$ referring to intrinsic), where the blue cloud and red-sequence in the colour-stellar mass relation appear clearly separated (since the $u$ filter is dominated by emission from massive and hence young stars and $r^{*}$ is dominated by the older population). \citet{Trayford15} showed that these colours are in agreement with various observational data for which dust corrections have been estimated (see e.g. \citealt{Schawinski14}). 

\begin{figure*} 
\begin{center}
\includegraphics[angle=0,width=0.85\textwidth]{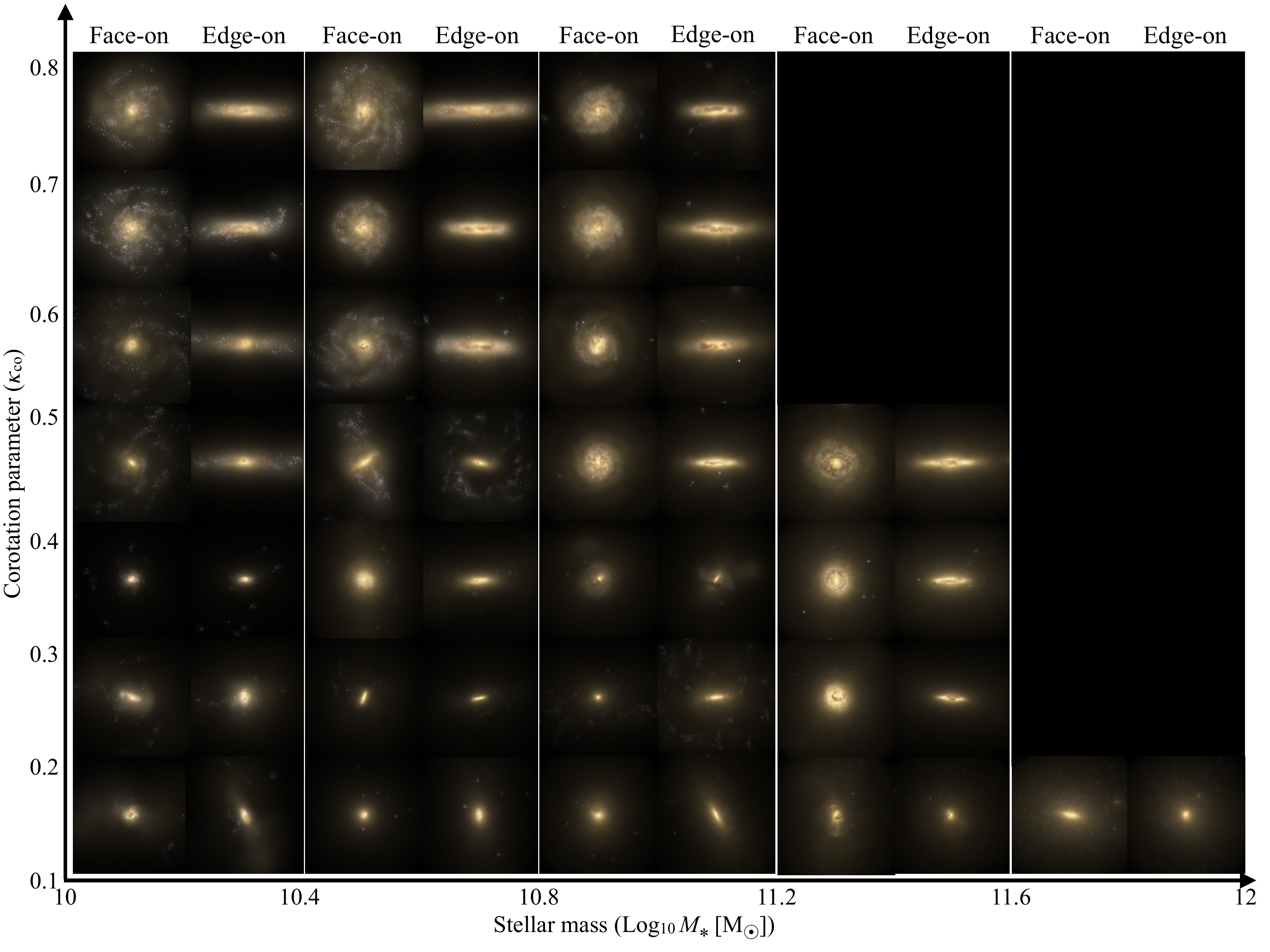}
\end{center}
\vspace{-0.5cm}
\caption{Example of central galaxies taken from the Ref-L100N1504 EAGLE simulation. The panels correspond to either face- or edge-on {\it{gri}}-composite images of side length 40 kpc. From left to right the figure shows galaxies with increasing mass, and from bottom to top galaxies with an increasing fraction of their stellar kinetic energy invested in ordered corotation. The empty bins indicate that there are no rotation-supported massive galaxies in EAGLE.}
\label{front_pic}
\end{figure*}

To quantify the morphology of a galaxy, we can use the fraction of kinetic energy invested in ordered rotation (\citealt{Sales10}),

$$\kappa_{\rm{rot}}=\frac{K_{\rm{rot}}}{K}=\frac{1}{K}\sum_{i}^{r<30\rm{kpc}}\frac{1}{2}\,m_i\left[L_{z,i}/(m_i\,R_i)\right]^{2},$$

\noindent where the sum is over all stellar particles within a spherical radius of 30  kpc centered on the minimum of the potential, $m_{i}$ is the mass of each stellar particle, $K(=\sum_{i}^{r<30\rm{kpc}}\frac{1}{2}m_iv_i^2)$ the total kinetic energy, $L_{z,i}$ the particle angular momentum along the direction of the total angular momentum of the stellar component of the galaxy ($\vec{L}$ with the velocity of the frame being the velocity of the stellar centre of mass, so that $L_{z,i}=L_i^{{\rm along}\,\vec{L}=\sum_{i}^{r<30\rm{kpc}}\vec{L_i}}$) and $R_{i}$ is the projected distance to the axis of rotation ($\vec{L}$).

Different from the literature, we calculate $K_{\rm{rot}}$ considering only star particles that follow the direction of rotation of the galaxy (i.e. with positive $L_i^{{\rm along}\,\vec{L}=\sum_{i}^{r<30\rm{kpc}}\vec{L_i}}$), and define $\kappa_{\rm{co}}$ (hereafter $\kappa$ {\it{corotating}}) as the fraction of kinetic energy invested in ordered corotation. Therefore $\kappa_{\rm{co}}\leq\kappa_{\rm{rot}}$. Fig.~\ref{color_morpho_distribution_1} shows a scatter plot of the relative difference between $\kappa_{\rm{co}}$ and $\kappa_{\rm{rot}}$ as a function of stellar mass. We can see that the median relative difference between $\kappa_{\rm{co}}$ and $\kappa_{\rm{rot}}$ (solid curve) increases from $\approx -10\%$ for $M_{*}<10^{10.5}\Msun$ to $-40\%$ for $M_{*}\sim 10^{11.5}\Msun$. We find that $\kappa_{\rm{co}}$ is a better measure of the importance of ordered rotation, since it does not consider star particles that are counterrotating and hence are likely part of the bulge. 

\section{Results}\label{color_morpho_sec}

\begin{figure*} 
\begin{center}
\includegraphics[angle=0,width=0.8\textwidth]{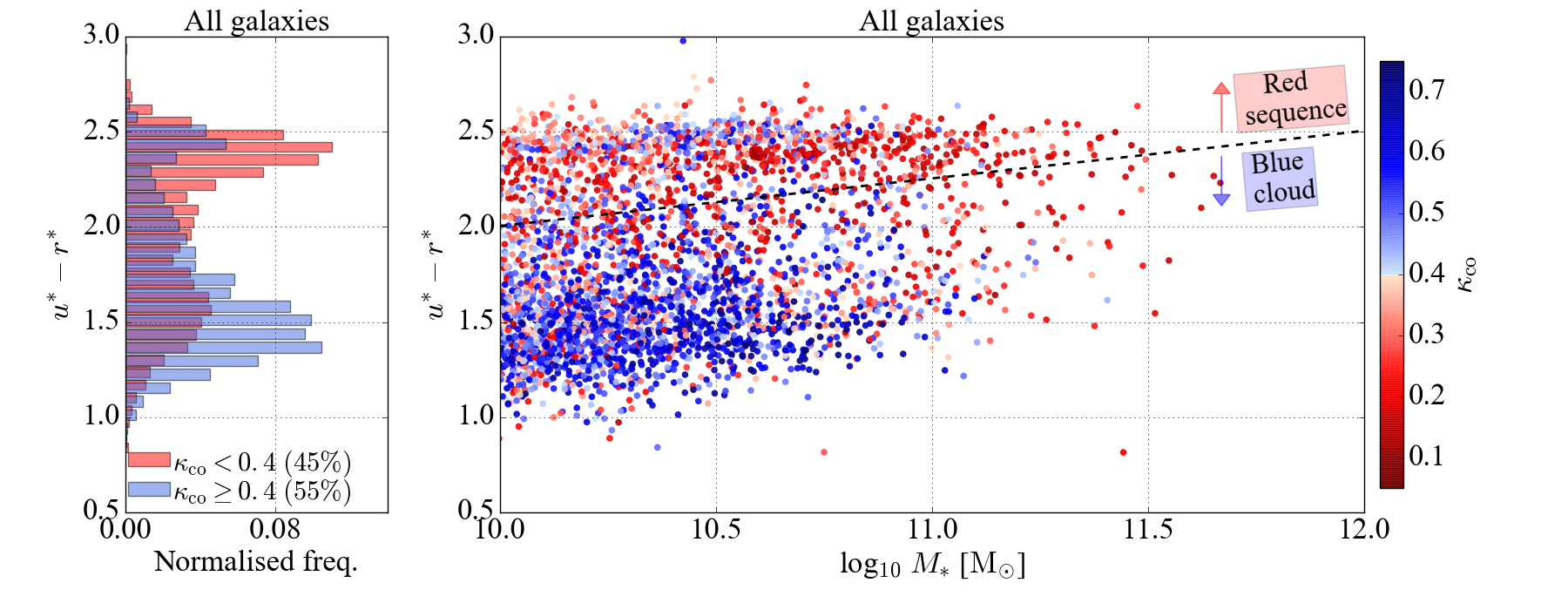}
\includegraphics[angle=0,width=0.49\textwidth]{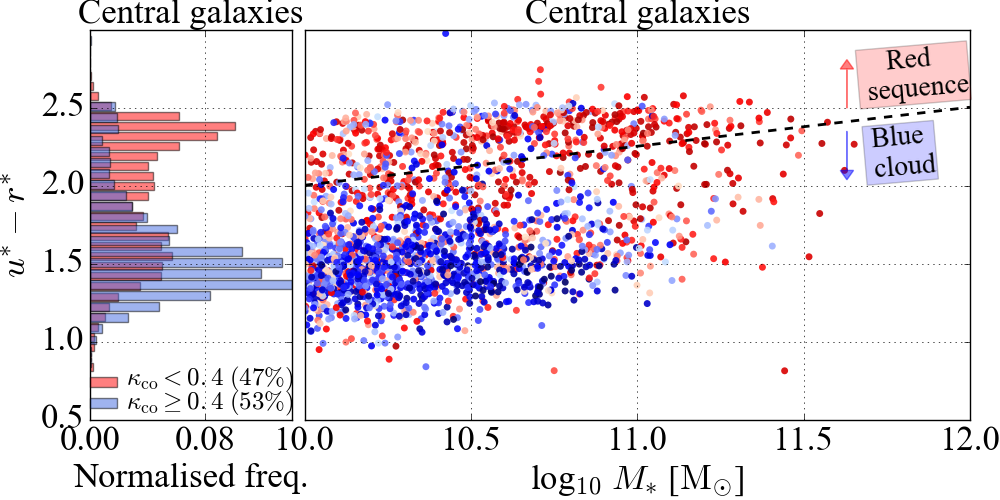}
\includegraphics[angle=0,width=0.49\textwidth]{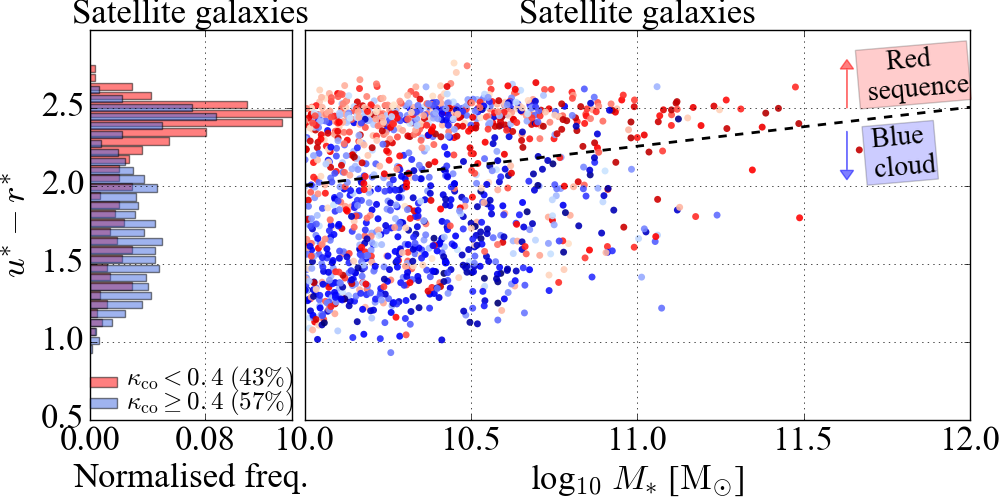}
\end{center}
\vspace{-0.3cm}
\caption{Intrinsic $u^{*}-r^{*}$ colour versus stellar mass, with individual galaxies plotted as points, coloured by $\kappa_{\rm{co}}$ according to the colour bar at the top right. The black dashed line indicates the adopted definitions of the red sequence and blue cloud. The top right, bottom left and bottom right panels show all galaxies, centrals and satellites, respectively. The left side of each panel, shows normalized histograms of the colour distribution of disc ($\kappa_{\rm{co}}\ge 0.4$, blue) and elliptical ($\kappa_{\rm{co}}< 0.4$, red) galaxies.}
\label{color_morpho_distribution_2}
\end{figure*}

Fig.~\ref{front_pic} shows images of 26 representative central galaxies distributed in bins of log$_{10}M_{*}$ and $\kappa_{\rm{co}}$. For each galaxy there is a face- and edge-on image generated using the SKIRT code (\citealt{Camps15}), with galaxev (\citealt{Bruzual03}) and Mappings III (\citealt{Groves08}) spectra to represent star particles and young HII regions respectively, as described by \citet{Trayford15}. A square field of view of 40 kpc on a side is used for observations in the SDSS gri bands (\citealt{Doi10}) to approximate SDSS colours. No artificial seeing is added to the images. These images are publicly available from the EAGLE database (\citealt{McAlpine16}).

Fig.~\ref{front_pic} shows from left to right galaxies with increasing stellar mass, and from bottom to top, galaxies with increasing $\kappa_{\rm{co}}$, i.e. a larger fraction of the stellar kinetic energy invested in ordered corotation. As $\kappa_{\rm{co}}$ increases, galaxies tend to become more disky and extended. Based on visual inspection of a larger number of galaxies we pick $\kappa_{\rm{co}}=0.4$ to separate galaxies that look disky from those that look elliptical. We will refer to galaxies with $\kappa_{\rm{co}}<0.4$ as ellipticals and galaxies with $\kappa_{\rm{co}}\ge 0.4$ as discs.

Fig.~\ref{color_morpho_distribution_2} shows scatter plots of $u^{*}-r^{*}$ colour versus stellar mass for all (top panels), central (bottom left panels) and satellite (bottom right panels) galaxies. Individual galaxies are coloured by $\kappa_{\rm{co}}$. The dashed line corresponds to $(u^{*}-r^{*})=0.25\log_{10}(M_{*}/\Msun)-0.495$ that crosses the green-valley as estimated by \citet{Schawinski14} and separates the red-sequence from the blue-cloud region in the colour-mass diagram. On the left of each plot we show normalized histograms of galaxy colours for elliptical (red bars) and disc (blue bars) galaxies. We find that from a total sample of 3562 galaxies, almost half of the $38\%$ that are satellite are red at $z=0$, and most of the central galaxies are blue. In addition, we find that from the entire population (both satellite and centrals) $45\%$ are elliptical and $55\%$ are discs. 

The panels highlight that morphology correlates with both colour and mass for central and satellite galaxies alike. There is a clear bimodality in colour with a red sequence mostly populated by elliptical galaxies (with median $\kappa_{\rm{co}}\lesssim0.3$) and a blue cloud mostly populated by disc galaxies (with median $\kappa_{\rm{co}}\gtrsim 0.45$). We find that among galaxies with $M_{*}\lesssim 10^{10.5}\Msun$ centrals are blue and disc-dominated, while satellite galaxies are more evenly distributed in colour and the satellite red sequence is more morphologically diverse. This is expected, since it has been shown that central and satellite galaxies do not evolve in colour, morphology and star-formation rate in a similar manner due to the strong environmental dependence of these properties at low redshift (e.g. \citealt{Kauffmann04,Peng10,Wetzel12,vandeVoort17}). Galaxies with $M_{*}>10^{11}\Msun$ are however nearly exclusively elliptical and tend to reside on the red sequence.

\begin{figure} 
\begin{center}
\includegraphics[width=0.44\textwidth]{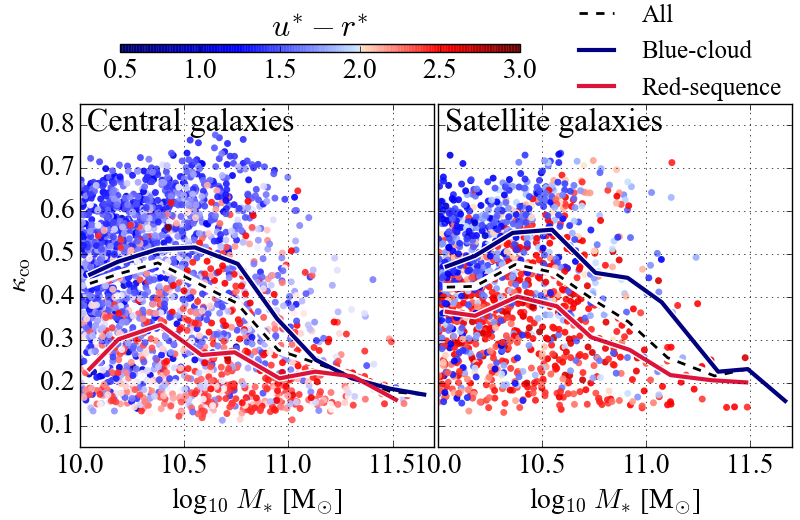}
\end{center}
\vspace{-0.4cm}
\caption{$\kappa_{\rm{co}}$ versus stellar mass for central (left) and satellite (right) galaxies. Individual galaxies are plotted as points, coloured by intrinsic $u^{*}-r^{*}$ according to the colour bar on the top. The solid lines show the median relations for blue-cloud and red-sequence galaxies. The black dashed lines show the median relation for the complete sample of centrals and satellites, respectively.}
\label{color_morpho_distribution_3}
\end{figure}

Fig.~\ref{color_morpho_distribution_3} shows scatter plots of $\kappa_{\rm{co}}$ as a function of stellar mass for centrals (left panel) and satellites (right panel). Individual galaxies are plotted as points, coloured by $u^{*}-r^{*}$. Solid lines correspond to the median $\kappa_{\rm{co}}$ for galaxies classified as blue-cloud and red-sequence. For blue-cloud galaxies the median $\kappa_{\rm{co}}$ peaks at $M_{*}\sim 10^{10.5}\Msun$. For higher masses $\kappa_{\rm{co}}$ declines with mass for all galaxy colours. While most central red-sequence galaxies are elliptical with median $\kappa_{\rm{co}}\approx 0.3$, satellite red-sequence galaxies less massive than $10^{10.7}\Msun$ have median $\kappa_{\rm{co}}\approx 0.4$ and so tend to be more disky.

\section{Conclusions}

We investigated the relation between kinematic galaxy morphology, intrinsic $u^{*}-r^{*}$ colour and stellar mass in the EAGLE simulation. We used the intrinsic colours from \citet{Trayford15} and measured $\kappa_{\rm{co}}$, the fraction of kinetic energy invested in ordered corotation, of 3562 galaxies at $z=0$ with stellar masses larger than $10^{10}\Msun$. In contrast to previous studies, we only considered star particles that follow the direction of rotation of the galaxy in the calculation of the kinetic energy in ordered rotation, in order not to count star particles that are likely part of the bulge. From visual inspection of {\it{gri}}-composite images of central galaxies we found that $\kappa_{\rm{co}}$ is a reasonable measure of visual morphology. We found that EAGLE produces a galaxy population whose morphology correlates with the colour bimodality for central and satellite galaxies alike. The red-sequence is mostly populated by elliptical galaxies and most blue-cloud galaxies are discs. With increasing mass, galaxies become redder and more elliptical. Satellite galaxies tend to be redder and for $M_{*}\lesssim 10^{10.5}\Msun$ the satellite red sequence is more morphologically diverse than for centrals. From these results we conclude that the observed connection between mass, intrinsic colour and morphology can arise naturally from galaxy formation models that reproduce the galaxy mass function and galaxy sizes. In future work we plan to investigate the origin of the morphological distribution of galaxies and its correlation with colour, as well as to perform like-for-like comparisons with photometric observations.

\section*{Acknowledgments}

This work used the DiRAC Data Centric system at Durham University, operated by the Institute for Computational Cosmology on behalf of the STFC DiRAC HPC Facility (www.dirac.ac.uk). This equipment was funded by BIS National E-infrastructure capital grant ST/K00042X/1, STFC capital grant ST/H008519/1, and STFC DiRAC Operations grant ST/K003267/1 and Durham University. DiRAC is part of the National E-Infrastructure. This work was supported by the European Research Council under the European Union's Seventh Framework Programme (FP7/2007-2013)/ERC Grant agreement 278594-GasAroundGalaxies and by the Netherlands Organisation for Scientific Research (NWO) through VICI grant 639.043.409. RAC is a Royal Society University Research Fellow. RGB acknowledges support from STFC (ST/L00075X/1).

\bibliography{biblio}
\bibliographystyle{mn2e}

\end{document}